\documentclass[aps,superscriptaddress,superbib,twocolumn,groupedaddress]{revtex4}

\usepackage[latin1]{inputenc}
\usepackage[T1]{fontenc}

\usepackage{amsmath,amssymb}
\usepackage{graphicx}
\graphicspath{{fig/}}

\usepackage{units}
\usepackage{journames}

\usepackage[final]{neel}

\usepackage{hyperref}
\hypersetup{pagebackref,breaklinks,citecolor=magenta,colorlinks,bookmarks,bookmarksopen,
  pdfauthor="Olivier Fruchart",pdfstartview=FitH}

\begin{document}

\renewcommand\topfraction{0.8}
\renewcommand\bottomfraction{0.7}
\renewcommand\floatpagefraction{0.7}

\def\Htc{H_{t\mathrm{,c}}}

\title{Dimensionality cross-over in magnetism: from domain walls~(2D) to vortices~(1D)}%

\author{A. Masseb\oe uf}
\affiliation{CEA-Grenoble, INAC/SP2M/LEMMA, 17 rue des Martyrs, Grenoble, France}%
\author{O. Fruchart}
\email[]{Olivier.Fruchart@grenoble.cnrs.fr}\affiliation{Institut N\'{E}EL, CNRS \& Universit\'{e}
Joseph Fourier -- BP166 -- F-38042 Grenoble Cedex 9 -- France}%
\author{J. C. Toussaint}
\affiliation{Institut N\'{E}EL, CNRS \& Universit\'{e}
Joseph Fourier -- BP166 -- F-38042 Grenoble Cedex 9 -- France}%
\affiliation{Institut National Polytechnique de Grenoble -- France}%
\author{E. Kritsikis}
\affiliation{Institut N\'{E}EL, CNRS \& Universit\'{e}
Joseph Fourier -- BP166 -- F-38042 Grenoble Cedex 9 -- France}%
\affiliation{Institut National Polytechnique de Grenoble -- France}%
\affiliation{SPINTEC, URA 2512, CEA/CNRS, CEA/Grenoble, 38054 Grenoble Cedex 9, France}%
\author{L. Buda-Prejbeanu}
\affiliation{SPINTEC, URA 2512, CEA/CNRS, CEA/Grenoble, 38054 Grenoble Cedex 9, France}%
\author{F. Cheynis}
\affiliation{Institut N\'{E}EL, CNRS \& Universit\'{e}
Joseph Fourier -- BP166 -- F-38042 Grenoble Cedex 9 -- France}%
\altaffiliation[Present address]{CNRS, Aix-Marseille Universit\'{e}, CINAM-UPR3118, Campus de Luminy, Case 913, F-13288 
Marseille Cedex 09, France.}
\author{P. Bayle-Guillemaud}
\affiliation{CEA-Grenoble, INAC/SP2M/LEMMA, 17 rue des Martyrs, Grenoble, France}%
\author{A. Marty}
\affiliation{CEA-Grenoble, INAC/SP2M/NM, 17 rue des Martyrs, Grenoble, France}%

\date{\today}

\begin{abstract}

Dimensionality cross-over is a classical topic in physics. Surprisingly it has not been searched in
micromagnetism, which deals with objects such as domain walls~(2D) and
vortices~(1D). We predict by simulation a second-order transition between these two objects, with the wall length as 
the
Landau parameter. This was confirmed experimentally based on micron-sized flux-closure dots.

\end{abstract}


\maketitle

\vskip 0.5in

\vskip 0.5in


Dimensionality cross-over is a rich topic in theoretical and experimental physics. It has been widely addressed in 
the frame of phase transition and critical exponents, \eg in magnetism\cite{bib-MER1966,bib-POU1999}. Beyond this 
microscopic level it is known that materials in an ordered state may be split in domains\cite{bib-HUB1998b}. In the 
study of magnetization configurations, a field known as
micromagnetism, objects have tentatively been classified according to their dimensionality. Magnetic domains are~3D, 
Domain walls~(DWs) are~2D, Bloch lines (\ie so-called either vortices or anti-vortices) are~1D, Bloch points 
are~0D\cite{bib-DOE1968,bib-ARR1997}. Each class may serve as a boundary to the class of immediately-greater 
dimensionality: DWs are found at domain boundaries, Bloch lines inside domain walls to separate areas with opposite 
winding\cite{bib-ARR1997,bib-HUB1998b}, and Bloch points separate two parts of a vortex with opposite 
polarities\cite{bib-DOE1968,bib-THI2003}. Beyond magnetism, the notions of DWs and vortices are shared by all states 
of matter ordered with a unidirectional order parameter, \ie characterized by a vector field $\vect n$ with $|\vect 
n|=1$. Liquid crystals in the nematic state have a uniaxial order parameter. A strict analogue of magnetic materials 
is the common case of slabs with anchoring conditions at both surfaces: upon application of a magnetic or electric 
field perpendicular to the easy axis of anchoring a breaking of symmetry occurs known as the Fredericks 
transition\cite{bib-ZOC1933}, transforming the order parameter in a unidirectional one. In this case both 
$\angledeg{180}$ domain walls and vortices occur\cite{bib-BRO1972,bib-deG1974}, whose dynamics, topology and 
annihilation are being studied \cite{bib-BLA2005}.

The study of magnetic DWs and vortices as objects that can be moved\cite{bib-GRO2003b} and 
modified\cite{bib-OKU2002,bib-VAN2006,bib-HER2007,bib-FRU2009b} in their inner structure is a timely topic, driven 
by proposals of their use in memory\cite{bib-ZHU2000} and logic\cite{bib-ALL2005} devices. Despite this and the 
established dimensional classification above-mentioned, surprisingly the possibility of a dimensionality cross-over 
between a DW and a vortex has not been addressed. Thus, beyond the aesthetics physical issue of dimensionality 
crossover, the knowledge of how a DW may switch reversibly to a magnetic vortex should have a great importance in 
understanding and controlling their static and dynamic features. This transition has not been described either in 
analogous cases such as liquid crystals.

\begin{figure}
  \begin{center}
  \includegraphics[width=86mm]{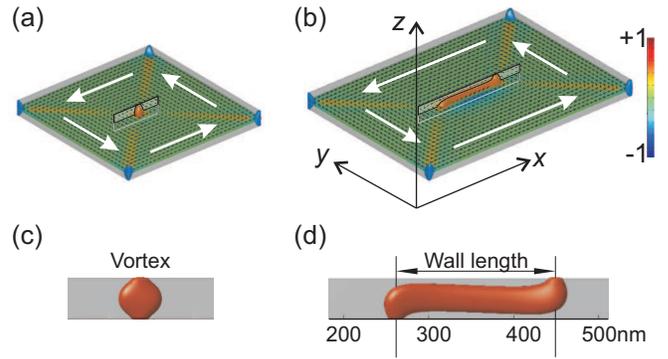}%
  \caption{\label{fig-square-rectangle}Simulated magnetization states in flat Fe(110) dots with size
     (a)~\lengthnm{$500\times500\times50$}: vortex state  and (b)~\lengthnm{$500\times750\times50$}: Landau state
     consisting of
     a Bloch wall separating two antiparallel domains. The color
     stands for the direction of magnetization along $z$, see right scale. In these open views the only
     parts displayed as volumes are those were $m_z$ is greater than~$0.5$. This highlights the central vortex or 
Bloch wall~(red) and the magnetization      areas close to the vertical edges of the prisms~(blue). At all 
other places the surface
     displays magnetization in the mid-height plane. (c-d)~Views in the $xz$ plane, corresponding to the framed 
areas in a-b, respectively.
     \dataref{Simuls-Calculs$\backslash$ Longueur de
     vortex$\backslash$
      Rapport-exploit-pave-droit-Lx500x40nm.doc}}
  \end{center}
\end{figure}

In this Letter we report on a dimensionality cross-over from a DW~(2D) to a vortex~(1D). Although exemplified in the 
particular case of magnetic materials, this effect should occur in any state of matter characterized with a 
unidirectional order parameter. It should be noted that a dimensionality cross-over was recently reported for the 
dynamics of motion of a domain wall along a stripe, whose behavior changes from two-dimensional pinning to 
one-dimensional pinning on structural defects when the width of the stripe is reduced below typically the distance 
between major pinning sites\cite{bib-KIM2009b}. This process however is very different, since it pertains to dynamic 
processes, and also is extrinsic because it relies on structural defects which depend on the material, method of 
deposition, and nanofabrication.

For our demonstration we considered epitaxial micron-sized magnetic dots in a flux-closure state. Depending on the 
dot geometry (size and aspect ratio), the flux may be closed around a vortex\cite{bib-OKU2002} or a DW of finite 
length\cite{bib-FRU2003c}. The use of an epitaxial material ensures that the results are not biased by 
microstructural pinning. Besides the dots display sharp and well-defined edges, so that their dimensions can be 
measured with accuracy. It is known that the topology of Bloch DW of finite length and of vortices is identical, the 
former being obtained from the latter by a continuous deformation\cite{bib-ARR1979b,bib-HER1999,bib-FRU2003c}. Thus 
the question of a transition  from a DW to a magnetic vortex arises naturally.
We first show by simulation that a wall tends to collapse into a vortex at a critical length of a few tens of 
nanometers. The transition is of second order, with the wall length as the order parameter. This is confirmed 
quantitatively by experiments based on micron-sized self-assembled epitaxial dots, both with the dot geometry and an 
external magnetic field as the driving parameter, showing the generality and robustness of the transition.

\begin{figure}
  \begin{center}
  \includegraphics[width=83mm]{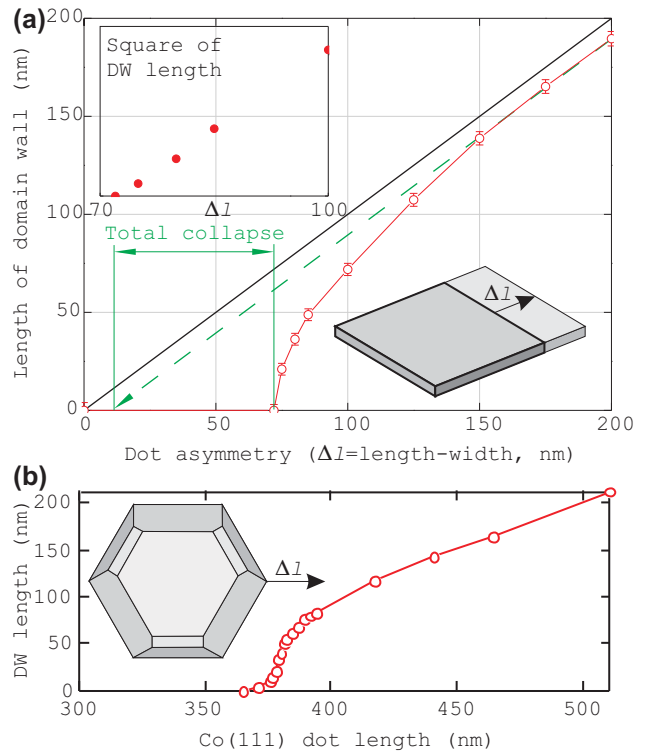}%
  \caption{\label{fig-second-order}(a)~Open symbols: length of the Bloch wall in the simulated Landau state
  in an Fe(110) dot of thickness \thicknm{50}, plotted versus the dot lateral asymmetry $\Delta l$ (length minus 
width, see inset sketch).
  Linear line with slope~$+1$: wall length in the simple geometrical Van den Berg model (black line). Dotted
  line: asymptotic extrapolation from long dot, whose intercept with the $x$ axis defines the total
  collapsed. Inset: squared length of Bloch wall, same $x$ axis. (b)~Similar simulations based on a Co(111) dot, 
here $\thicknm{50}$-thick. The inset shows the detailed facetted shape of these dots\cite{bib-FRUunpub}.}
  \end{center}
\end{figure}

Let us describe our methods. Micromagnetic simulations of prisms were performed using GL\_FFT, a finite-differences 
code\cite{bib-TOU2002}. The cell size was $\thicknm{3.91\times3.91\times3.13}$ or lower and the parameters for bulk 
Fe were used with an Fe(110) orientation\cite{bib-FRU2009b}. Simulations of trigonal fcc Co(111) dots were performed 
using Feellgood, a finite-elements code\cite{bib-feellgood-tbp}. The mean cell size was $\lengthnm{2.6}$. Both codes 
have been custom-developed and are based on the temporal integration of the Landau-Lifshitz-Gilbert equation.
The experimental systems consist of self-assembled micron-size Co(111)\cite{bib-FRUunpub} and 
Fe(110)\cite{bib-FRU2009b} dots, epitaxially-grown under ultra-high vacuum using pulsed-laser 
deposition\cite{bib-FRU2007}. These were grown on a single-crystalline \thicknm{10}-thick W buffer layer deposited 
on sapphire $(11\overline20)$ wafers, and capped with a \thicknm{5}-thick Au layer to prevent oxidation. The wafer 
was then thinned by mechanical polishing and ion milling.
Lorentz microscopy was performed in the Fresnel mode using a JEOL 3010 microscope equipped with a GATAN imaging 
filter. In this mode  DWs~(resp. vortices) are highlighted as dark or bright lines~(resp. dots) depending on the 
chirality of magnetization curling around the DW/vortex\cite{bib-CHA1984}. The image are formed with a dedicated 
mini-lens, while an axial magnetic field can be added using the conventional objective lens.

We first present the results of simulation. As a textbook case we consider flat prismatic  dots with fixed 
height-over-width ratio $0.2$ and thickness $\thicknm{50}$ and above. The prismatic shape and use of finite 
differences ensure high accuracy results for the description of the phase transition. The length, taken along the 
in-plane $\mathrm{Fe}[001]$ direction, is varied from $1$ to $1.5$ with respect to the width.
As expected for elongated dots of such thickness\cite{bib-ARR1979b,bib-HER1999,bib-FRU2003c} a Landau state occurs, 
displaying two main longitudinal domains separated by a Bloch DW\bracketsubfigref{fig-square-rectangle}{b,d}. The DW 
displays perpendicular magnetization in its core, while it is terminated at each surface by an area with in-plane 
magnetization, the \textsl{N\'{e}el caps}\cite{bib-HUB1969}. At each end of the DW the magnetic flux escapes through a 
surface vortex. We define the length of the DW as the distance between the projections into the film plane of the 
locii of these two vortices\bracketsubfigref{fig-square-rectangle}{d}.
From this definition a vortex is a DW with zero length, such as found \eg for a dot with a square 
base\bracketsubfigref{fig-square-rectangle}{a,c}.
Series of simulations of the equilibrium state for variable dot length were performed. At each stage the 
magnetization map is stretched or compressed along the length to serve as a crude input for the map of the next 
value of length, for which the equilibrium state is again calculated. The series was performed once with rising 
length, then again with decreasing length back to the square base. This yielded identical results, ruling out the 
possibility of metastable configurations biasing the results.
To avoid discretization artifacts the number of cells was kept constant for all simulations of a dot of given 
height. The length of each cell was varied progressively to fit the dot length.
The dependence of the DW length with the dot length is shown on \subfigref{fig-second-order}a. For largely elongated 
dots the wall length increases linearly with slope~1. In this regime the two surface vortices are sufficiently apart 
one from another to have a negligible interaction. Their position is essentially determined by the minimization of 
the energy of the triangular closure domain along the two short sides of the dot.
On the reverse in the low-length regime the DW length decreases faster than slope~1, so that the vortex
state is reached before the dot has a square base. We define the \textsl{collapsed} length as the difference between 
the length of dot upon the collapse and the asymptotic linear variation of wall length for an elongated 
dot\bracketsubfigref{fig-second-order}a. Plotting the square of the DW length versus the dot length reveals a linear 
variation. The cross-over is therefore Landau-like, \ie of second order. Such transitions are associated with a 
breaking of symmetry, which in the present case is whether the top surface vortex shifts towards $+x$ or $-x$. The 
results are qualitatively similar for other thicknesses. \subfigref{fig-second-order}b shows the results of similar 
finite-element simulations applied to trigonal Co(111) dots also \lengthnm{50}-thick, which display the very same 
physics. This suggests that the cross-over is a general phenomenon, independent from the exact shape of the system.

\begin{figure}
  \begin{center}
  \includegraphics[width=83mm]{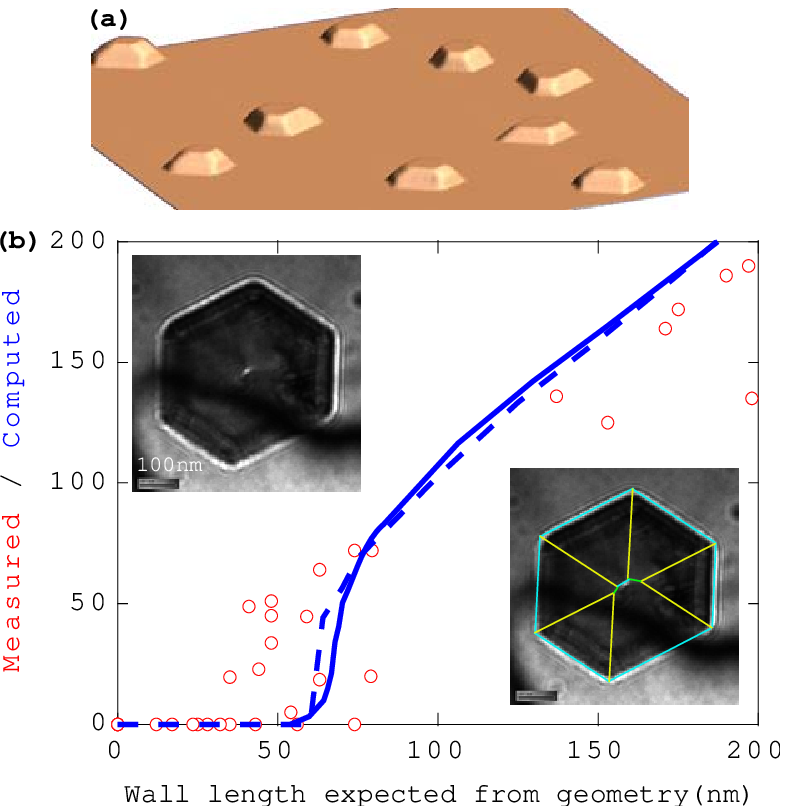}%
  \caption{\label{fig-co-dots}\dataref{(a) TP ENSPG 2008 (b) manips Lorentz 2009-09-12}(a)~True Z-scale 3D view of a 
$\thickmicron{6\times6}$ AFM image of self-assembled Co(111) dots (b)~Open symbols: DW length measured in Co(111) 
dots plotted versus the length expected from the geometrical Van den Berg construction. The predictions from 
simulations for the estimated thickness of the dot~$\thicknm{116}$ are superimposed without adjustment, as guide 
to the eye, both for Fe(110~(dotted line) and Co(111)~(full line). Insets: typical Co dot displaying a 
vortex~(upper left) along with the associated construction predicting a DW~(lower right, central blue line).}
  \end{center}
\end{figure}

These predictions have been confirmed experimentally. We first consider self-assembled face-centered cubic Co(111) 
dots. These have a trigonal symmetry reflecting their crystalline structure, with a base close to a regular 
hexagon\bracketsubfigref{fig-co-dots}b\cite{bib-FRUunpub}. Whereas previous studies on such systems were dealing 
with very thin dots thus found in a nearly single-domain state\cite{bib-LI2002}, the thickness of our dots is in the 
range $\thicknm{50-200}$ inducing flux-closure states around a DW or vortex.
These dots are perfectly suited for our needs because owing to the natural spread of shape occurring in 
self-assembly we can study the length of DWs as a function of the dot aspect ratio, by a statistical investigation 
of an assembly of dots over the same sample. For each dot we measured the experimental DW length, and computed the 
expected DW length predicted by the simple Van den Berg geometrical construction. This construction is relevant for 
vanishing thickness and infinite dimensions\cite{bib-VAN1984}, and equals the dot asymmetry used in the simulations 
so that a direct comparison with the data of \figref{fig-second-order} is possible. \subfigref{fig-co-dots}b 
summarizes this analysis, performed over more than 30 dots. The collected results are quantitatively consistent with 
the simulation predictions. The experimental spread of points may be attributed first to errors in the measurement 
of both the DW length and dot dimensions, second to the spread of dot thickness as the collapsing length slightly 
depends on the thickness. Despite this spread, it shall be noticed that only vortices are observed when the expected 
length lies below $\thicknm{40}$. This cannot be attributed to an experimental limitation to identify short DWs, as 
many DWs with length below $\thicknm{40}$ have been measured. These however all lie for expected wall length above 
$\thicknm{40}$. These correlations lie above statistical fluctuations, which unambiguously demonstrates the collapse 
of DWs towards vortices in a quantitative agreement with simulations.

\begin{figure}
  \begin{center}
  \includegraphics[width=86.16mm]{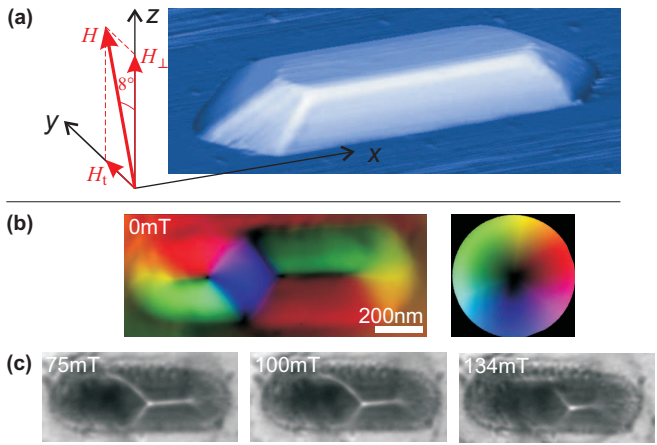}%
  \caption{\label{fig-fe(110)}(a)~True Z-scale 3D AFM view of a typical Fe(110) dots, along with the geometry of the 
applied field. (b)~map of the in-plane magnetization of a similar dot found in the diamond state, reconstructed 
from Fresnel images taken at different defocusing. (c)~Series of Fresnel images of the same dot taken under a 
magnetic field applied tilted towards $y$. The labels indicate the value of the $y$ component of the field.}
  \end{center}
\end{figure}

To demonstrate the generality and robustness of the vortex-DW transformation we now consider an external field as 
the driving parameter for the transition. In this case we use Fe(110) dots, which by their crystallographic nature 
are elongated \cite{bib-FRU2007}\bracketsubfigref{fig-fe(110)}a. Upon applying the saturating field with a tilt 
angle of a few degrees with respect to the normal to the plane, and the in-plane component oriented along its short 
length, the dot can be prepared in a diamond state, \ie consisting of two flux-closure parts with opposite 
chiralities\bracketsubfigref{fig-fe(110)}b. It happens that the application of a tilted field of moderate magnitude 
affects the length of the DWs\bracketsubfigref{fig-fe(110)}c in a continuous way. The length of the DW to the right 
of the dot decreases with increasing field. When its reaches $\lengthnm{21\pm3}$, which is comparable to the 
collapse length mentioned above, a stochastic switching was observed in real time between a Bloch wall and a vortex 
state with a characteristic time constant of $\unit[100]{ms}$~(see supplementary material). We ascribe this as a 
confirmation of the attraction of the top and bottom surface vortices, being the driving force for the transition.

To conclude we addressed the dimensionality cross-over of a magnetic domain wall~(DW, 2D) into a magnetic 
vortex~(1D). Simulations and experiments agree quantitatively that DWs collapse into vortices at a critical length 
of a few tens of nanometers, which reveals a short-ranged attractive force between the two ends of a DW. Beyond 
physics aesthetics, our investigation should prove useful when analyzing the increasing number of experiments 
dealing with the behavior of domain walls and vortices under the effect of pulsed magnetic fields or spin-polarized 
currents, which undergo complex variations of shape and length during their dynamics. This includes the case of \eg 
the vortex state, where a domain wall dynamically replaces the vortex\cite{bib-FISCHER-review}, or the 
multiplication of vortices or
transformation of the type of domain wall in magnetic stripes\cite{bib-NAK2003,bib-KLA2006}.

\section*{References}


\end{document}